\documentclass{appolb}
\usepackage{graphicx}

\usepackage{amsmath} 
\usepackage{xcolor}
\usepackage[numbers]{natbib}
\usepackage{hyperref}

\definecolor{darkgreen}{rgb}{0,0.35,0.15}
\definecolor{Green4}{rgb}{0.,0.376,0.}
\definecolor{Gray}{rgb}{0.2,0.2,0.2}
\definecolor{Gray2}{rgb}{0.35,0.35,0.35}
\definecolor{Orange}{rgb}{0.9,0.4,0.}
\definecolor{Black}{rgb}{0.,0.,0.}
\definecolor{Blue}{rgb}{0.,0.,1.}
\definecolor{Green}{rgb}{0.,1.,0.}
\definecolor{Cyan}{rgb}{0.,1.,1.}
\definecolor{Red}{rgb}{1.,0.,0.}
\definecolor{Magenta}{rgb}{1.,0.,1.}
\definecolor{Yellow}{rgb}{1.,1.,0.}
\definecolor{White}{rgb}{1.,1.,1.}
\definecolor{Blue4}{rgb}{0.,0.,0.5625}
\definecolor{Blue3}{rgb}{0.,0.,0.6875}
\definecolor{Blue2}{rgb}{0.,0.,0.8125}
\definecolor{LtBlue}{rgb}{0.52734375,0.8046875,1.}
\definecolor{Green3}{rgb}{0.,0.6875,0.}
\definecolor{Green2}{rgb}{0.,0.8125,0.}
\definecolor{Cyan4}{rgb}{0.,0.5625,0.5625}
\definecolor{Cyan3}{rgb}{0.,0.6875,0.6875}
\definecolor{Cyan2}{rgb}{0.,0.8125,0.8125}
\definecolor{Red4}{rgb}{0.5625,0.,0.}
\definecolor{Red3}{rgb}{0.6875,0.,0.}
\definecolor{Red2}{rgb}{0.8125,0.,0.}
\definecolor{Magenta4}{rgb}{0.5625,0.,0.5625}
\definecolor{Magenta3}{rgb}{0.6875,0.,0.6875}
\definecolor{Magenta2}{rgb}{0.8125,0.,0.8125}
\definecolor{Brown4}{rgb}{0.5,0.1875,0.}
\definecolor{Brown3}{rgb}{0.625,0.25,0.}
\definecolor{Brown2}{rgb}{0.75,0.375,0.}
\definecolor{Pink4}{rgb}{1.,0.5,0.5}
\definecolor{Pink3}{rgb}{1.,0.625,0.625}
\definecolor{Pink2}{rgb}{1.,0.75,0.75}
\definecolor{Pink}{rgb}{1.,0.875,0.875}
\definecolor{Gold}{rgb}{1.,0.83984375,0.}

\definecolor{background}{cmyk}{0,0,0.3,0}
\definecolor{dgreen}{rgb}{0,.4,0}
\definecolor{plum}{rgb}{.7 .2 .7}
\definecolor{darkgreen}{rgb}{.0 .6 .0}
\definecolor{peru}{rgb}{.80 .52 .25}

\begin{document}
\title{Center regions as a solution to the Gribov problem\\of the center vortex model
\thanks{Presented at Excited QCD 2020 by Rudolf Golubich}
}
\author{Rudolf Golubich and Manfried Faber 
\address{Atominstitut, Technische Universit\"at Wien}
}
\maketitle
\begin{abstract}
The center vortex model, capable of explaining confinement and chiral symmetry breaking, has been plagued by the lattice equivalent of Gribov copies: different maxima of the gauge functional lead to different predictions of the string tension. It is possible to resolve this problem using center regions, loops evaluating to center elements, as guide for the gauge fixing procedure. The success of this approach was already shown, but the algorithms came with an arbitrary free parameter. In recent development this parameter has been fixed, even improving the results.
\end{abstract}

\section{Introduction} 
\begin{figure}[h!]
\begin{center}
\includegraphics[height=6.5cm]{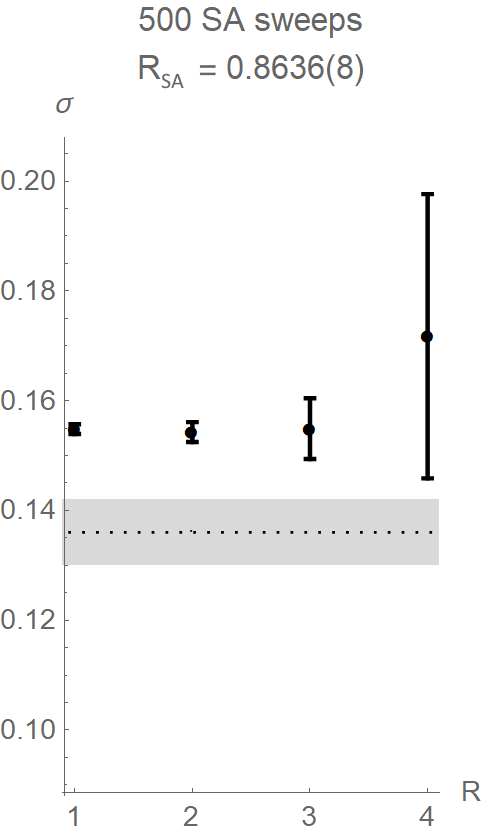}\hspace{5mm}
\includegraphics[height=6.5cm]{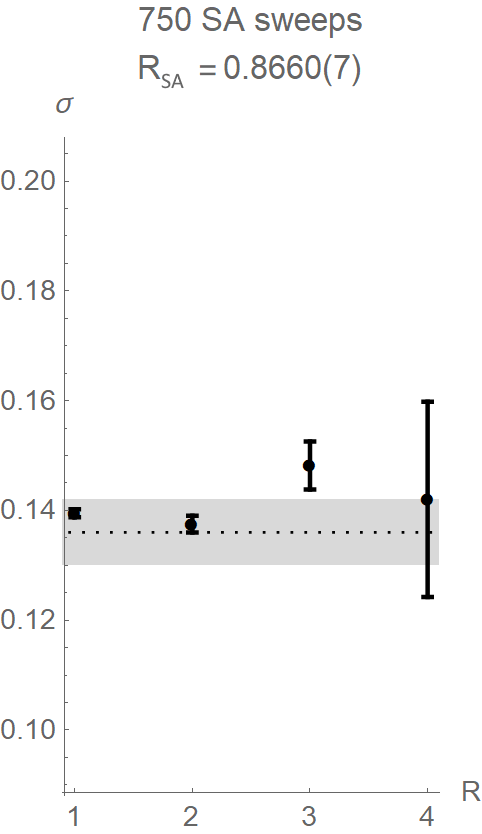}\hspace{5mm}
\includegraphics[height=6.5cm]{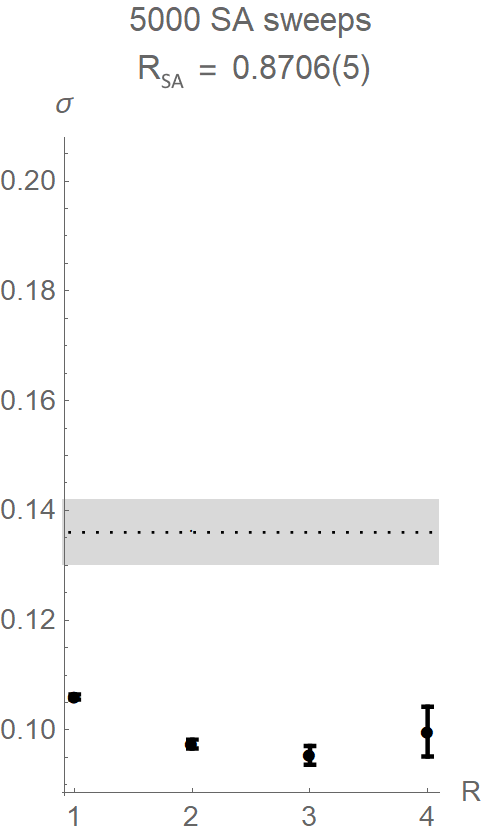}
\includegraphics[width=8cm]{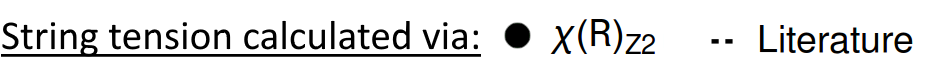}
\caption{\label{under} The string tension was calculated with 300 SU(2) Wilson configurations at $\beta=2.3$, in lattices of size $12^4$ (left), $12^4$ (middle) and $14^4$ (right). The literature value is based on \cite{Bali1995}. With increased number of simulated annealing steps the value of the gauge functional improves, but the string tension is lost.}
\end{center}
\end{figure}

The \textit{center vortex model}~\cite{THOOFT,CORNWALL} is based upon the center symmetry of the action in lattice quantum chromodynamics. It describes the properties of the vacuum by percolating vortices, that is, closed, quantized magnetic flux lines of finite thickness, condensing in the vacuum. It can explain the behaviour of Wilson and Polyakov loops \cite{DelDebbio:1998luz}, broken scale invariance \cite{Langfeld:1997jx} and 
Chiral symmetry breaking \cite{Faber:2017alm}, although it still lacks a closed mathematical formulation. To detect vortices in an SU(2) lattice, the gauge is fixed to \textit{Direct Maximal Center Gauge} using simulated annealing: generating random gauge matrices $\Omega(x)$, one numerically looks for maximizing the functional 
\begin{equation}
 R_{SA}^{2}= \sum_x \sum_\mu \mid \text{Tr}[\Omega(x) U_{\mu}(x) \Omega^{\dagger}(x+e_{\mu}] \mid^2,
\end{equation}
with $U_{\mu}(x)$ being an element of SU(2) corresponding to the link pointing in direction $\mu$ at lattice point $x$. A projection to the center degrees of freedom follows
\begin{equation}
U_{\mu}(x) \rightarrow Z_{\mu}(x) = \text{sign } \text{Tr}[\Omega(x) U_{\mu}(x) \Omega^{\dagger}(x+e_{\mu}].
\end{equation}
The string tension $\sigma$ is estimated using Creutz ratios 
\begin{equation}
\chi(R,T)=-\ln\frac{\langle W(R+1,T+1) \rangle \; \langle W(R,T) \rangle}{\langle W(R,T+1) \rangle \; \langle W(R+1,T) \rangle},
\end{equation}
with $W(R,T)$ being a Wilson loop of size $R\times T$ evaluated in the center projected configuration after gauge fixing. From the asymptotic relation $\langle W(R,T) \rangle = e^{- \sigma \; R \; T - 2 \; \mu \; (R+T) + C}$ with sufficiently high $R$ and $T$ follows $\chi=\sigma$.
\\
Simulated annealing can lead to different local maxima of the gauge functional $R_{SA}$ with differing physical properties: an improvement in the value of the gauge functional can lead to an underestimation of the string tension $\sigma$, see \cite{Bornyakov:2002sb, Faber:2001hq} and Figure \ref{under}, showing calculations for an SU(2) lattice.

By enforcing restrictions on the gauge matrices $\Omega(x)$ we can prevent this loss of the string tension: only such transfomations are allowed, that result in non-trivial center regions \cite{goluFab2020} projecting onto -1. Previously published versions of the respective algorithms came with a free parameter \cite{RudolfGolubich65935846}. Now we present a parameter free version, further improving the results.

\section{Detecting non-trivial center regions}
\begin{figure}[h!]
\begin{center}
1)~\parbox[c]{3.4cm}{
\includegraphics[width=2.5cm]{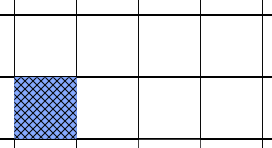}
} 2)~\parbox[c]{3.4cm}{
\includegraphics[width=2.5cm]{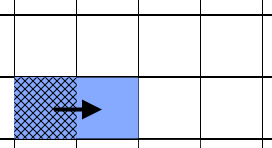}
} 3)~\parbox[c]{3.4cm}{
\includegraphics[width=2.5cm]{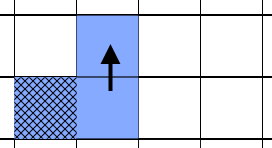}
}\\[1mm]
\parbox{12cm}{\footnotesize{\underline{Steps 1-3}: Starting with a plaquette that neither belongs to a before identified center region, nor was taken as origin for growing a region, it is tested, which enlargement by a neighboring plaquette brings the new region nearest to a center element.}} 
\noindent 4)~\parbox[c]{3.4cm}{
\includegraphics[width=2.5cm]{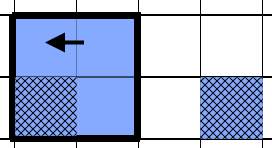}
} 5)~\parbox[c]{3.4cm}{
\includegraphics[width=2.5cm]{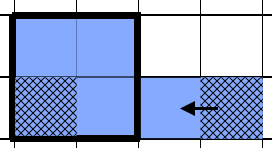}
} 6)~\parbox[c]{3.4cm}{
\includegraphics[width=2.5cm]{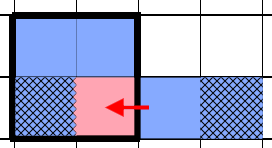}
}\\[1mm]
\parbox{12cm}{\footnotesize{\underline{Steps 4-6}: If no enlargement leads to further improvement, a new enlargement procedure is started with another plaquette. During enlargement the new region could grow into an existing one. The following steps describe the collision handling:}}
\noindent 7a)~\parbox[c]{4.7cm}{
\includegraphics[width=2.5cm]{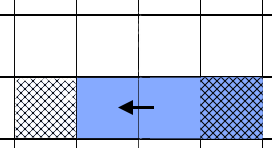}
} 7b)~\parbox[c]{4.7cm}{
\includegraphics[width=2.5cm]{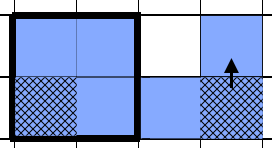}
}\\[1mm]
\parbox{5.5cm}{\footnotesize{\underline{Step 7a}: The evaluation of the growing region is nearer to a non-trivial center element than the evaluation of the old region: Delete the old region, only keeping the mark on its starting plaquette and allow growing.}
}\hspace{0.5cm}\parbox{5.5cm}{\footnotesize{\underline{Step 7b}: The growing region deviates more from a non-trivial center element than the existing one: try other enlargements.}}
\caption[Algorithm for region detection]{\label{CrAlgoOverall}The algorithm for detecting center regions repeats these procedures until every plaquette either belongs to an identified region or has been taken once as starting plaquette for growing a region. The arrow marks the direction of enlargement, plaquettes belonging to a region are colored, plaquettes already used as origin are shaded.}
\end{center}
\end{figure}

\begin{figure}[h!]
\begin{center}
\includegraphics[width=10cm]{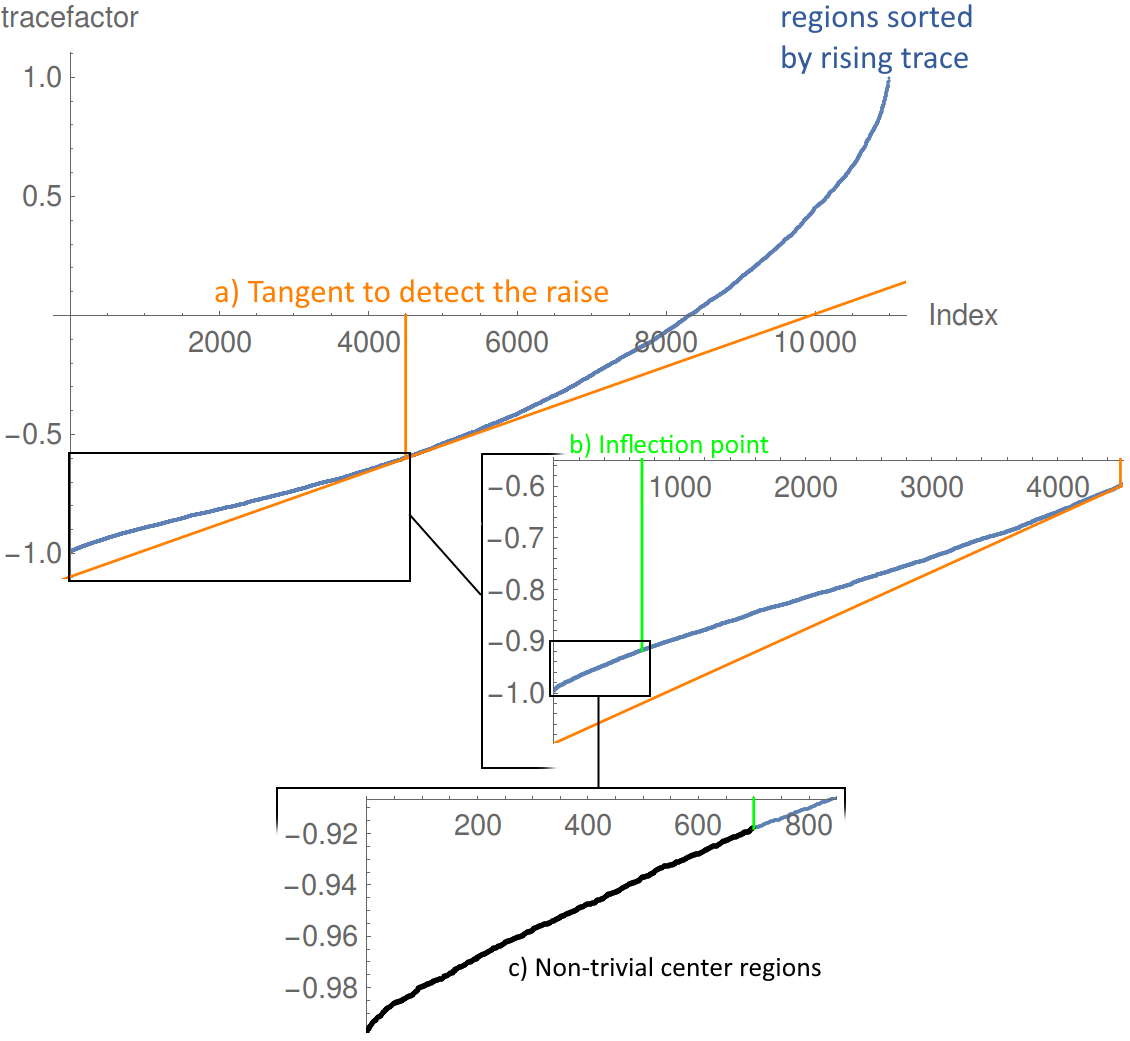}
\end{center}
\caption{Non-trivial center regions are selected within a list of detected regions sorted by rising trace. In this list a second rise in trace is identified by fitting a tangent (a). Within the interval from lowest trace to this tangent point, the lowest single inflection point is identified (b). All regions below this inflection point are taken for further usage (c).}
\label{crSelection}
\end{figure}

\noindent After sorting the plaquettes of a configuration by rising traces, the list is processed taking the plaquettes as origin for an enlargement procedure as shown in Figure \ref{CrAlgoOverall}.
This procedure aims to push the evaluation of the growing region as close towards a non-trivial center element as possible. Not all resulting regions evaluate sufficiently near to a non-trivial center element, hence a further selection is necessary. The identified regions are sorted by rising trace. Within this sorted list an inflection point before a stronger rise in the values of the traces is identified, see Figure \ref{crSelection}. The interval, the inflection point is searched in, is defined by a tangent through the point given by 1.1 times the value of the lowest trace at index 0. This is followed by identifying the lowest possible single inflection point using a method based on second difference quotients.  
All center regions within the list below this inflection point are taken into account during the gauge fixing procedure by simulated annealing. This procedure is modified if non-trivial center regions appear or vanish: appearence is always enforced, disappearence always forbidden.

\section{Results}
\begin{figure}[h!]
\begin{center}
\includegraphics[width=3.8cm]{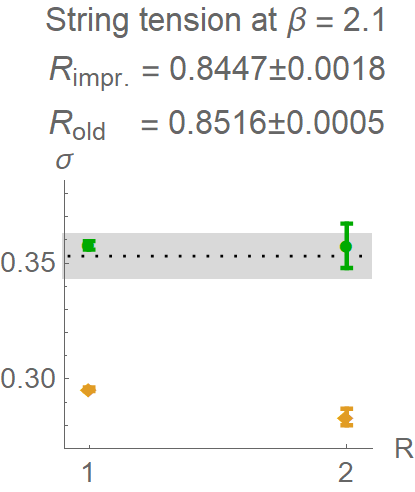}\hspace{3mm}
\includegraphics[width=3.8cm]{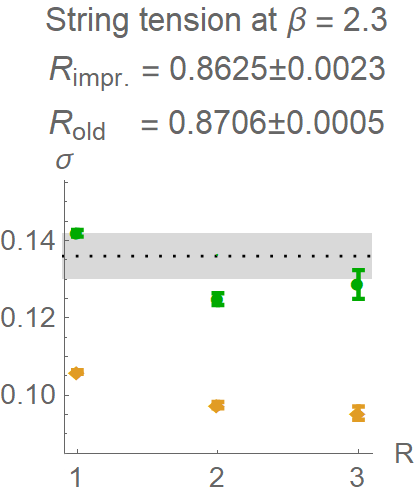}\hspace{3mm}
\includegraphics[width=3.8cm]{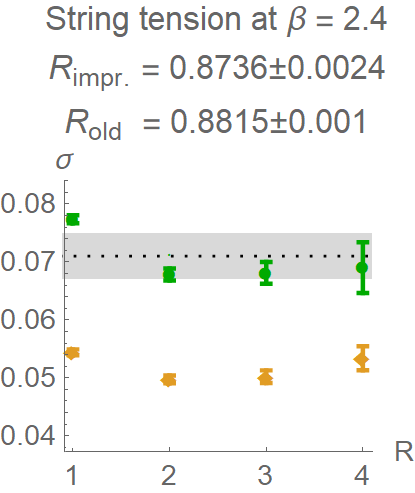}
\includegraphics[width=5.2cm]{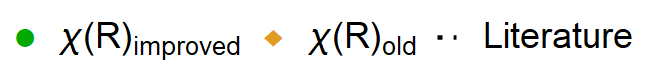}
\caption{\label{compare}We compare the original simulated annealing with our improved procedures after 5000 simulated annealing steps for 300 SU(2) Wilson configurations in lattices of size $12^4$ (left), $14^4$ (middle) and $12^4$ (right). The literature value is based on \cite{Bali1995}. Our improvements clearly recover the string tension.}
\end{center}
\end{figure}

\begin{figure}[h!]
\begin{center}
\includegraphics[height=5cm]{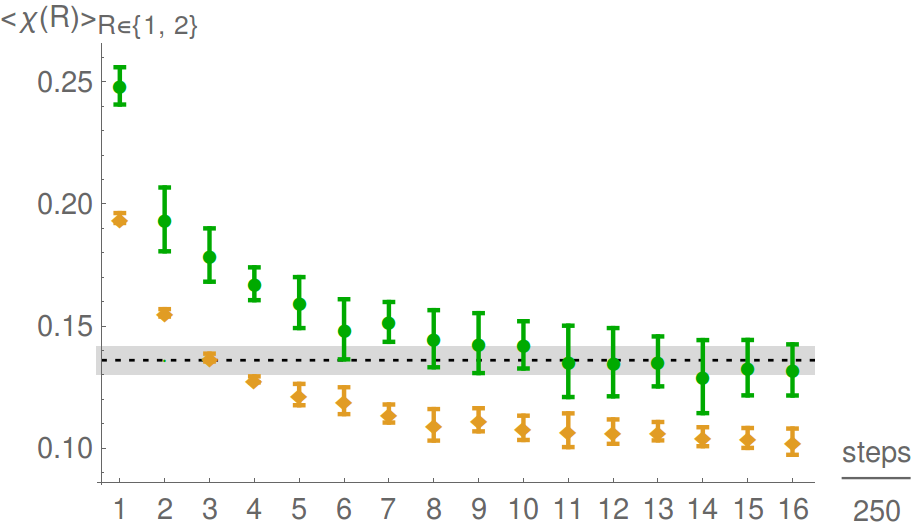}
\includegraphics[width=6cm]{img/LegendII.png}
\end{center}
\caption{\label{simstep} With rising number of simulated annealing steps the original procedure leads to an underestimation of the string tension, while our improvements stay on literature value beginning with 2500 steps. The data was calculated in a lattice of size $14^4$ for $\beta=2.3$. The literature value is based on \cite{Bali1995}.}
\end{figure}

The original simulated annealing procedure has the problem to result in an underestimation of the string tension. By preserving non-trivial center regions the string tension can be recovered at the cost of a small reduction in the value of the gauge functional and an increase in the size of the errorbars, see Figure \ref{compare}.
The dependence on the number of simulated annealing steps, shown in Figure \ref{simstep}, indicates that the improved procedure stays on literature value after a sufficient number of steps. 

\bibliographystyle{unsrtnat}
\bibliography{literature}

\begin{thebibliography}{10}
\providecommand{\natexlab}[1]{#1}
\providecommand{\url}[1]{\texttt{#1}}
\expandafter\ifx\csname urlstyle\endcsname\relax
  \providecommand{\doi}[1]{doi: #1}\else
  \providecommand{\doi}{doi: \begingroup \urlstyle{rm}\Url}\fi

\bibitem['t~Hooft(1978)]{THOOFT}
G.~'t~Hooft.
\newblock On the phase transition towards permanent quark confinement.
\newblock \emph{Nuclear Physics B}, 138\penalty0 (1):\penalty0 1 -- 25, 1978.
\newblock ISSN 0550-3213.
\newblock \doi{https://doi.org/10.1016/0550-3213(78)90153-0}.
\newblock URL
  \url{http://www.sciencedirect.com/science/article/pii/0550321378901530}.

\bibitem[Cornwall(1979)]{CORNWALL}
John~M. Cornwall.
\newblock Quark confinement and vortices in massive gauge-invariant qcd.
\newblock \emph{Nuclear Physics B}, 157\penalty0 (3):\penalty0 392 -- 412,
  1979.
\newblock ISSN 0550-3213.
\newblock \doi{https://doi.org/10.1016/0550-3213(79)90111-1}.
\newblock URL
  \url{http://www.sciencedirect.com/science/article/pii/0550321379901111}.

\bibitem[Del~Debbio et~al.(1998)Del~Debbio, Faber, Giedt, Greensite, and
  Olejnik]{DelDebbio:1998luz}
L.~Del~Debbio, Manfried Faber, J.~Giedt, J.~Greensite, and S.~Olejnik.
\newblock {Detection of center vortices in the lattice Yang-Mills vacuum}.
\newblock \emph{Phys. Rev.}, D58:\penalty0 094501, 1998.
\newblock \doi{10.1103/PhysRevD.58.094501}.

\bibitem[Langfeld et~al.(1998)Langfeld, Reinhardt, and
  Tennert]{Langfeld:1997jx}
Kurt Langfeld, Hugo Reinhardt, and Oliver Tennert.
\newblock {Confinement and scaling of the vortex vacuum of SU(2) lattice gauge
  theory}.
\newblock \emph{Phys. Lett.}, B419:\penalty0 317--321, 1998.
\newblock \doi{10.1016/S0370-2693(97)01435-4}.

\bibitem[Faber and H\"ollwieser(2017)]{Faber:2017alm}
Manfried Faber and Roman H\"ollwieser.
\newblock {Chiral symmetry breaking on the lattice}.
\newblock \emph{Prog. Part. Nucl. Phys.}, 97:\penalty0 312--355, 2017.
\newblock \doi{10.1016/j.ppnp.2017.08.001}.

\bibitem[Bornyakov et~al.(2002)Bornyakov, Komarov, Polikarpov, and
  Veselov]{Bornyakov:2002sb}
V.~G. Bornyakov, D.~A. Komarov, M.~I. Polikarpov, and A.~I. Veselov.
\newblock {P vortices, nexuses and effects of Gribov copies in the center
  gauges}.
\newblock In \emph{{Quantum chromodynamics and color confinement. Proceedings,
  International Symposium, Confinement 2000, Osaka, Japan, March 7-10, 2000}},
  pages 133--140, 2002.

\bibitem[Faber et~al.(2001)Faber, Greensite, and Olejnik]{Faber:2001hq}
Manfried Faber, Jeff Greensite, and Stefan Olejnik.
\newblock {Remarks on the Gribov problem in direct maximal center gauge}.
\newblock \emph{Phys. Rev.}, D64:\penalty0 034511, 2001.
\newblock \doi{10.1103/PhysRevD.64.034511}.

\bibitem[Bali et~al.(1995)Bali, Schlichter, and Schilling]{Bali1995}
Gunnar~S. Bali, Christoph Schlichter, and Klaus Schilling.
\newblock Observing long color flux tubes in su(2) lattice gauge theory.
\newblock \emph{Physical Review D}, 51\penalty0 (9):\penalty0 5165–5198, May
  1995.
\newblock ISSN 0556-2821.
\newblock \doi{10.1103/physrevd.51.5165}.
\newblock URL \url{http://dx.doi.org/10.1103/PhysRevD.51.5165}.

\bibitem[Golubich and Faber(2020)]{goluFab2020}
Rudolf Golubich and Manfried Faber.
\newblock {The Road to Solving the Gribov Problem of the Center Vortex Model in
  Quantum Chromodynamics}.
\newblock \emph{Acta Physica Polonica B Proceedings Supplement}, 13:\penalty0
  59--65, 2020.
\newblock \doi{10.5506/APhysPolBSupp.13.59}.

\bibitem[Golubich and Faber(2019)]{RudolfGolubich65935846}
Rudolf Golubich and Manfried Faber.
\newblock Improving center vortex detection by usage of center regions as
  guidance for the direct maximal center gauge.
\newblock \emph{Particles}, 2019.
\newblock \doi{10.3390/particles2040030}.
\newblock URL \url{http://doi.org/10.3390/particles2040030}.

\end{thebibliography}

\end{document}